# Unexpected mass acquisition of Dirac fermions at the quantum phase transition of a topological insulator

T. Sato[1], Kouji Segawa[2], K. Kosaka[1], S. Souma[3], K. Nakayama[1], K. Eto[2], T. Minami[2], Yoichi Ando[2], & T. Takahashi[1,3]

[1]*Department of Physics, Tohoku University, Sendai 980-8578, Japan*

[2]*Institute of Scientific and Industrial Research, Osaka University, Ibaraki, Osaka 567-0047, Japan*

[3]*WPI Research Center, Advanced Institute for Materials Research, Tohoku University, Sendai 980-8577, Japan*

**The three-dimensional (3D) topological insulator is a novel quantum state of matter where an insulating bulk hosts a linearly-dispersing surface state, which can be viewed as a sea of massless Dirac fermions protected by the time-reversal symmetry (TRS). Breaking the TRS by a magnetic order leads to the opening of a gap in the surface state[1] and consequently the Dirac fermions become massive. It has been proposed theoretically that such a mass acquisition is necessary for realizing novel topological phenomena[2,3], but achieving a sufficiently large mass is an experimental challenge. Here we report an unexpected discovery that the surface Dirac fermions in a solid-solution system $TlBi(S_{1-x}Se_x)_2$ acquires a mass *without* explicitly breaking the TRS. We found that this system goes through a quantum phase transition from the topological to the non-topological phase, and by tracing the evolution of the electronic states using the angle-resolved photoemission, we observed that the massless Dirac state in $TlBiSe_2$ switches to a massive state before it disappears in the**

**non-topological phase. This result suggests the existence of a condensed-matter version of the "Higgs mechanism" where particles acquire a mass through spontaneous symmetry breaking.**

Whether a band insulator is topological or not is determined by the parity of the valence-band wave function, which is described by the $Z_2$ topological invariant. Strong spin-orbit coupling can lead to an inversion of the character of valence- and conduction-band wave functions, resulting in an odd $Z_2$ invariant that characterizes the topological insulator[4,5]. All known topological insulators[6-14] are based on this band-inversion mechanism[4,5,15-18], but a successive evolution of the electronic state across the quantum phase transition (QPT) from trivial to topological has not been well studied in 3D topological insulators due to the lack of suitable materials. $TlBi(S_{1-x}Se_x)_2$ is therefore the first system where one can investigate the 3D topological QPT[19]. The advantage of this system is that it always maintains the same crystal structure (Fig. 1a) irrespective of the S/Se ratio. The low-energy, ultrahigh-resolution angle-resolved photoemission spectroscopy (ARPES) that has recently become available is particularly suited to trace such a QPT in great detail.

The bulk band structures of the two end members, $TlBiSe_2$ and $TlBiS_2$, are shown in Fig. 1b, where one can see several common features such as the prominent hole-like band at the binding energy $E_B$ of 0.5–1 eV and a weaker intensity at the Fermi level ($E_F$), both are centered at the $\bar{\Gamma}$ point (Brillouin-zone center). These features correspond to the top of the bulk valence band (VB) and the bottom of the bulk conduction band (CB), respectively, demonstrating that both $TlBiSe_2$ and $TlBiS_2$ samples are originally an insulator with a band gap of 0.3-0.4 eV, but electron carriers are doped in the naturally-grown crystals[10-12]. Besides the wider VB width in $TlBiS_2$ that is naturally expected from its smaller lattice constant, the VB structures in the two systems are very similar (Fig. 1c).

A critical difference in the electronic states of the two compounds is recognized by looking at the band dispersion in the vicinity of $E_F$ around the $\overline{\Gamma}$ point (Fig. 1d). An "X"-shaped surface band which traverses the bulk band gap is well recognized in TlBiSe$_2$ (ref. 10), whereas such a surface state is completely absent in TlBiS$_2$. This indicates the topologically distinct nature of the two, despite the similar overall bulk-band structure. One can thus conclude that the band parity is inverted in TlBiSe$_2$ while it is not in TlBiS$_2$.

What will happen if we mix the topologically non-trivial TlBiSe$_2$ and trivial TlBiS$_2$ phases? One natural consequence of such an alloying would be that the bulk band gap closes due to the switching of the parity of the VB wave function at a certain Se content $x_c$ marking a topological QPT, across which the massless Dirac-cone surface band appears (vanishes) once the system enters into the topological (non-topological) phase. The surface band in the topological phase would keep the Kramers degeneracy at the Dirac point and retain the massless character, as long as the alloying disorder does not break the TRS. It turns out that the electronic-structure evolution in TlBi(S$_{1-x}$Se$_x$)$_2$ indeed presents the topological QPT, but it bears a feature that is totally unexpected.

Figure 2a shows the near-$E_F$ ARPES intensity around the zone center for a series of $x$ values including TlBiSe$_2$ ($x = 1.0$) and TlBiS$_2$ ($x = 0.0$). One can immediately see that the surface state is seen for $x \geq 0.6$, whereas it is absent for $x \leq 0.4$ (see also Supplementary Information), which points to the topological QPT occurring at $x_c \approx 0.5$. In fact, the bulk band gap estimated from our data approaches zero on both sides of the QPT (see Fig. 3d), suggesting that a band inversion takes place across the QPT, in accordance with the natural expectation and also with a recent ARPES work independently done on

TlBi($S_{1-x}Se_x$)$_2$ (ref. 19).

The unexpected physics manifests itself at the Dirac point. The bright intensity peak at ~0.4 eV at $x = 1.0$ is no longer visible at $x = 0.9$ and is markedly suppressed at $x = 0.6$, suggesting that the Kramers degeneracy is lifted upon S substitution while the surface state is still present. In fact, a closer look at the energy-distribution curves (EDCs) in Fig. 2b finds that the originally X-shaped surface band at $x = 1.0$, where the EDC at the $\bar{\Gamma}$ point is well fitted by a slightly asymmetric Lorenzian, splits into lower and upper branches at $x = 0.9$ with a finite energy gap at the $\bar{\Gamma}$ point. Further substitution of S results in the reduction and the broadening of the surface-band intensity (see EDCs for $x = 0.6$), but the energy position of the surface band can be still traced, as illustrated in the second-derivative intensity plots in Fig. 2c. The surface-state nature of this band was confirmed by its stationary nature of the energy position with respect to the photon energy (Supplementary Information), so this band evidently represents the massive Dirac fermions on the surface. Because of the gapped nature, this phase may not be called topological in the strict sense, but the massive Dirac fermions are obviously of topological origin, suggesting that the bulk bands are kept inverted. On the other hand, the photon-energy dependence of the ARPES spectra for $x = 0.4$ signifies the absence of the surface state in contrast to the clear signature of it for $x = 0.6$ (Supplementary Information). The disappearance of the surface state and the very narrow bulk gap at $x = 0.4$ (which can be inferred in the Supplementary Fig. S2) point to the topological QPT located between $x = 0.6$ and 0.4. Interestingly, the surface band gap, called here the Dirac gap, grows upon decreasing $x$ (less than 0.1 eV at $x = 0.9$ and 0.8, and larger than 0.1 eV at $x = 0.6$), indicating that the S content is closely related to the magnitude of the Dirac gap. We also found that the magnitude of the Dirac gap does not diminish with increasing temperature (Supplementary Information), which speaks against a magnetic-order origin of the gap. We note that in a recent independent work[19], the Dirac cone was reported to remain gapless for $x > 0.5$, in contrast to the gapped surface states for $0.6 \leq x \leq 0.9$ observed here. This discrepancy may be due to the difference in the energy resolution (15 meV in ref. 19, as opposed to 2-4 meV in the present experiment).

One may wonder if the bulk band gap really closes at $x_c \approx 0.5$. If the band gap never closes, both the samples for $x = 1.0$ and 0.0 should be in the same topological phase. Apparently, this is inconsistent with our ARPES data in Fig. 1**d**. According to the fundamental principle of the topological band theory, the QPT *must* always be accompanied by a band gap closing. Hence, based on our data, the band gap closing must be happening either at $0.4 < x < 0.6$ or at $0.9 < x < 1.0$ (at which the Dirac gap starts to open). Taking into account the gradual reduction of the band-gap size on approaching $x = 0.5$, it is most sensible to conclude that the inevitable gap closing is taking place at $x = 0.5$.

One may also question if the observed Dirac gap might be an artifact of an inhomogeneous S distribution in the sample. To address this question, we employed the electron-probe microanalysis (EPMA) on the surface and found that our crystals are exceedingly homogeneous (Supplementary Information). The persistently-narrow x-ray diffraction peaks as well as a systematic change of the lattice constants further corroborate this conclusion (Supplementary Information).

To quantify the magnitude of the Dirac gap, we use the theoretical surface-band dispersion to account for the finite mass term[20] (which was originally proposed to explain the $Bi_2Se_3$ ultrathin-film data[21]) and numerically simulate the experimentally-obtained surface band dispersion near the $\bar{\Gamma}$ point, although the origin of the mass term in $TlBi(S_{1-x}Se_x)_2$ is not clear at the moment. As shown in Fig. 3a, the simulated curves reasonably reproduce the experimental data, and the obtained Dirac gap for $x$ = 0.9, 0.8 and 0.6 are 50±10, 70±10, and 130±20 meV, respectively. The evolution of the massive Dirac cone is schematically illustrated in 3D images of the band dispersions in Fig. 3b. We have confirmed that the obtained Dirac-gap sizes are highly reproducible by measuring more than 5 samples for each composition and also by varying the incident photon energy. Taking into account that all the elements contained in $TlBi(S_{1-x}Se_x)_2$ are nonmagnetic and also that the sample shows no obvious magnetic order (Supplementary Information), our result strongly suggests that the substitution of Se with S in $TlBi(S_{1-x}Se_x)_2$ leads to an unconventional mass acquisition of the surface Dirac fermions *without* explicitly breaking the TRS.

Based on the present ARPES results, one may draw the electronic phase diagram of $TlBi(S_{1-x}Se_x)_2$ as shown in Fig. 3d. The *massless* Dirac topological phase is achieved only near $x$ = 1.0. Once a small amount of S is substituted for Se, the Dirac gap opens and it grows almost linearly as a function of the S content, 1-$x$. Such a *massive* Dirac phase is present until the topological QPT occurs at $x_c \approx 0.5$, where the bulk gap closes and the band parity is interchanged.

The mass acquisition of the Dirac fermions indicates that the Kramers degeneracy is lifted, which means that the TRS must be broken on the surface. Given that there is no explicit TRS breaking, the only possibility is that a *spontaneous* symmetry breaking takes place upon the S substitution, which is reminiscent of the Higgs mechanism in particle physics. Therefore, $TlBi(S_{1-x}Se_x)_2$ may serve as a model system to bridge the condensed-matter physics and particle physics. The exact mechanism of the mass acquisition is not clear at the moment, but an interesting possibility is that it originates from some exotic many-body effects that can lead to an electronic order, although a simple mechanism like the spin-density wave does not seem to be relevant (Supplementary information). When the top and bottom surface states coherently couple and hybridize, the Dirac gap can open[21], but the sufficiently large thickness (> 10 µm) of our samples precludes this origin. Another possibility is that critical fluctuations associated with the QPT are responsible for the mass acquisition, but it is too early to speculate along this line. From the application point of view, the Dirac gap can be much larger than that of magnetically-doped topological insulator $Bi_2Se_3$ (ref. 1) and is tunable with the S/Se ratio, making the $TlBi(S_{1-x}Se_x)_2$ system a prime candidate for device applications that require a gapped surface state.

**METHODS**

High-quality single crystals of $TlBi(S_{1-x}Se_x)_2$ were grown by a modified Bridgman method (see Supplementary Information for details). X-ray diffraction measurement indicated the monotonic shrinkage of *a*- and *c*-axis lengths upon substitution of S for Se, without any apparent change in the relative atomic position with respect to the unit cell. ARPES measurements were performed using VG-SCIENTA SES2002 and MBS-A1 spectrometers with high-flux He and Xe discharge lamps and a toroidal / spherical grating monochromator at Tohoku University. The He Iα ($h\nu$ = 21.218 eV) line and one of the Xe I ($h\nu$ = 8.437 eV) lines[22] were used to excite photoelectrons. Samples were cleaved *in-situ* along the (111) crystal plane in an ultrahigh vacuum of $5\times10^{-11}$ Torr. The energy resolutions for the measurement of the VB and near-$E_F$ regions are set at 15 and 2-4 meV, respectively. The angular resolution was 0.2° corresponding to the *k* resolution of 0.007 and 0.004 Å$^{-1}$ for the He Iα and Xe I photons, respectively. The Fermi level of the samples was referenced to that of a gold film evaporated onto the sample holder. A shiny mirror-like surface was obtained after cleaving samples, confirming its high quality.

lamp for bulk-sensitive high-resolution photoemission spectroscopy. Rev. Sci. Instrum. **78**, 123104 (2007).

**Acknowledgements**

We thank N. Nagaosa for valuable discussions. We also thank H. Guo, K. Sugawara, M. Komatsu, T. Arakane, and A. Takayama for their assistance in the ARPES experiment, and S. Sasaki for the analysis using EPMA. This work was supported by JSPS (KAKENHI 19674002 and NEXT Program), JST-CREST, MEXT of Japan (Innovative Area "Topological Quantum Phenomena"), AFOSR (AOARD 10-4103), and KEK-PF (Proposal number: 2009S2-005 and 2010G507).

**Author Contributions**

T.S., K.K., S.S., K.N., and T.T. performed ARPES measurements. K.S. K.E, T.M. and Y.A. carried out the growth of the single crystals and their characterizations. T.S., K.S. and Y.A. conceived the experiments and wrote the manuscript.

**Competing Interests Statement**

We have no competing financial interests.

**Correspondence** and requests for materials should be addressed to T.S. (e-mail: t-sato@arpes.phys.tohoku.ac.jp) or Y.A. (e-mail: y_ando@sanken.osaka-u.ac.jp).

**FIGURE LEGENDS**

**FIG. 1: Comparison of the valence-band structure between TlBiSe$_2$ and TlBiS$_2$.** **a,** Crystal structure of TlBi(S$_{1-x}$Se$_x$)$_2$. **b,** Valence-band ARPES intensity along the $\bar{\Gamma}$- $\bar{K}$- $\bar{M}$ direction of the surface Brillouin zone for TlBiSe$_2$ and TlBiS$_2$ plotted as a function of wave vector and binding energy measured with the He Iα resonance line ($h\nu$ = 21.218 eV) at $T$ = 30 K. **c**, Direct comparison of the valence-band dispersions between TlBiSe$_2$ and TlBiS$_2$. Energy position of the bands are determined by tracing the peak position of the second-derivatives of the ARPES spectra. Several branches of dispersive bands in both compounds are found at $E_B$ higher than 1 eV, and they are attributed to the hybridized states of Tl 6$p$, Bi 6$p$ and Se 4$p$ (S 3$p$) orbitals[16-18]. Clearly, the bands in TlBiS$_2$ are shifted toward higher $E_B$ with respect to those in TlBiSe$_2$, and such an energy shift is more enhanced in deeper-lying bands, indicating the relative expansion of the VB width in TlBiS$_2$. **d**, Comparison of near-$E_F$ ARPES intensity around the $\bar{\Gamma}$ point between TlBiSe$_2$ and TlBiS$_2$ measured with the Xe I resonance line ($h\nu$ = 8.437 eV). Absence of the surface band in TlBiS$_2$ was also confirmed by varying photon energy and light polarization in synchrotron radiation.

**FIG. 2: Mass acquisition of surface Dirac fermions in TlBi(S$_{1-x}$Se$_x$)$_2$.** **a**, Near-$E_F$ ARPES intensity around the $\bar{\Gamma}$ point as a function of wave vector and binding energy in TlBi(S$_{1-x}$Se$_x$)$_2$ for a series of Se concentrations, measured with the Xe I line with the same experimental geometry. White dotted lines for $x \leq 0.4$ represent the top of VB and the bottom of CB to highlight the band gap; because of the $k_z$ dispersion of the bulk bands, the true gap is likely to be smaller than these estimates, which is reflected in the error bars in Fig. 3d. **b**, Energy distribution curves (EDCs) of the data shown in **a**. **c**,

Second-derivative plots of the ARPES intensity for $x$ = 1.0, 0.9, 0.8, and 0.6.

**FIG. 3: Massive Dirac fermions and the electronic phase diagram across the topological quantum phase transition.** **a**, Numerical fittings of the ARPES-determined surface band dispersions (open circles) using the theoretical band dispersion which incorporates the finite mass term[20]. Note that the Dirac energy for $x$ = 0.9 is shifted downward by 0.015 eV for clarity. **b**, Schematic 3D image of the surface Dirac cone and its energy-gap evolution at low sulfur concentrations. **c**, ARPES spectra around the VB and CB edges for $x$ = 1.0 and 0.6 which are used for the band-gap determination for $x$ > 0.5 as was done in ref. 10; the photon energies of the synchrotron radiation were chosen to optimally probe the VB top and the CB bottom respectively[10]. **d**, Electronic phase diagram of $TlBi(S_{1-x}Se_x)_2$ showing the surface Dirac gap (green symbols) and the bulk band gap (red symbols), together with a schematic picture of band evolution (top) summarizing the present ARPES experiment. Error bars for the Dirac gap energy correspond to the experimental uncertainty in determining the energy position of the band dispersion at the $\bar{\Gamma}$ point in Fig. 3**a**, whereas those for the bulk band gap originate from the experimental uncertainty in determining the energy difference between the top of the valence band and the bottom of the conduction band in Figs. 2 and 3**c**.

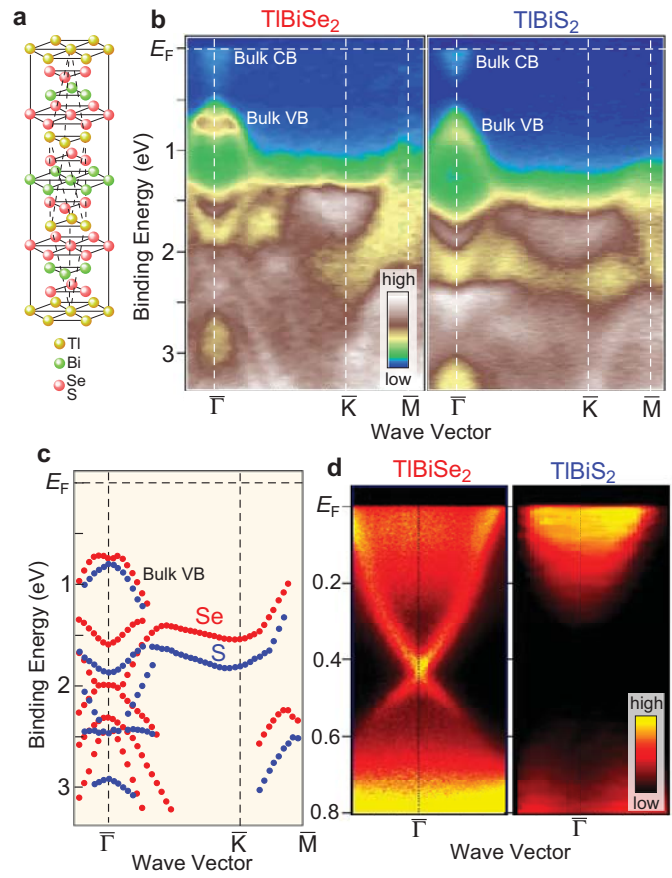

Fig. 1 Sato

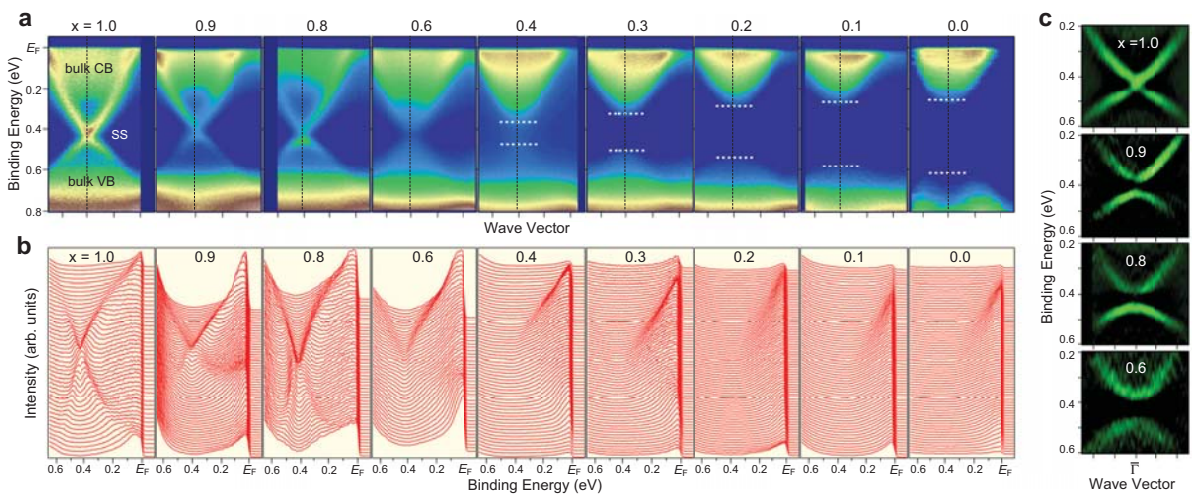

Fig. 2 Sato

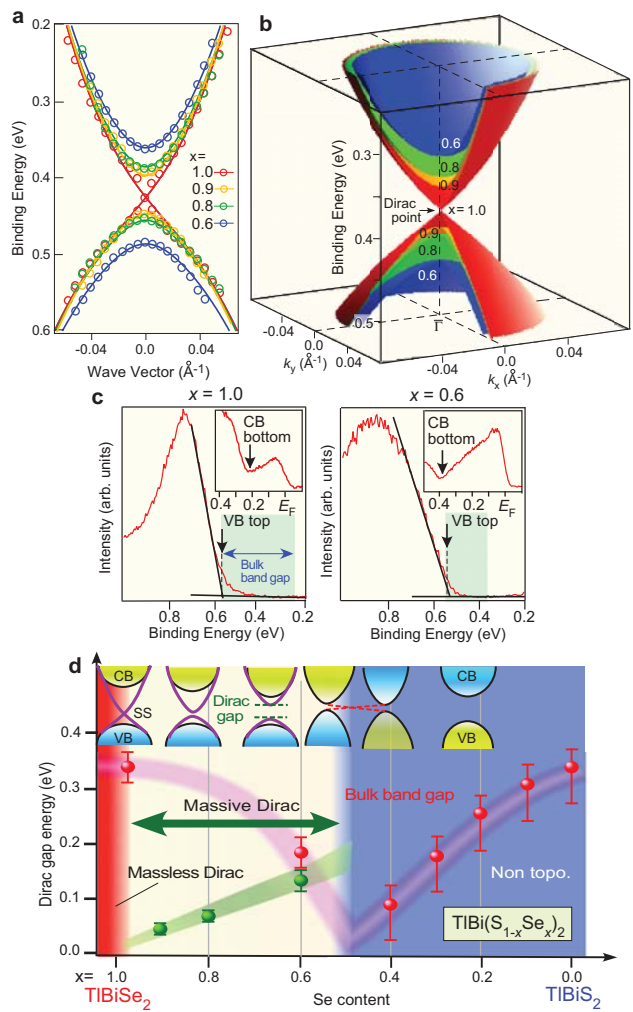

Fig. 3 Sato

**Supplementary information for "Unexpected mass acquisition of Dirac fermions at the quantum phase transition of a topological insulator"**

T. Sato[1], Kouji Segawa[2], K. Kosaka[1], S. Souma[3], K. Nakayama[1], K. Eto[2], T. Minami[2], Yoichi Ando[2], & T. Takahashi[1,3]

[1]Department of Physics, Tohoku University, Sendai 980-8578, Japan
[2]Institute of Scientific and Industrial Research, Osaka University, Ibaraki, Osaka 567-0047, Japan
[3]WPI Research Center, Advanced Institute for Materials Research,
Tohoku University, Sendai 980-8577, Japan

1. Preparation of the Single Crystals

Single crystals of a series of TlBi($S_{1-x}Se_x$)$_2$ were grown by a modified Bridgman technique, in which the temperature of the furnace was slowly lowered in 4 days while the melt was placed at a position where a temperature gradient of about 10 K was present. The optimum growth temperature varied with $x$: For example, it was 1095 - 975 K for $x$ = 0.0, while it was 1035 - 915 K for $x$ = 1.0. The purities of the starting materials were 99.9999% for Bi, Se and S, and 99.999% for Tl.

2. Identification of the Surface and Bulk Bands

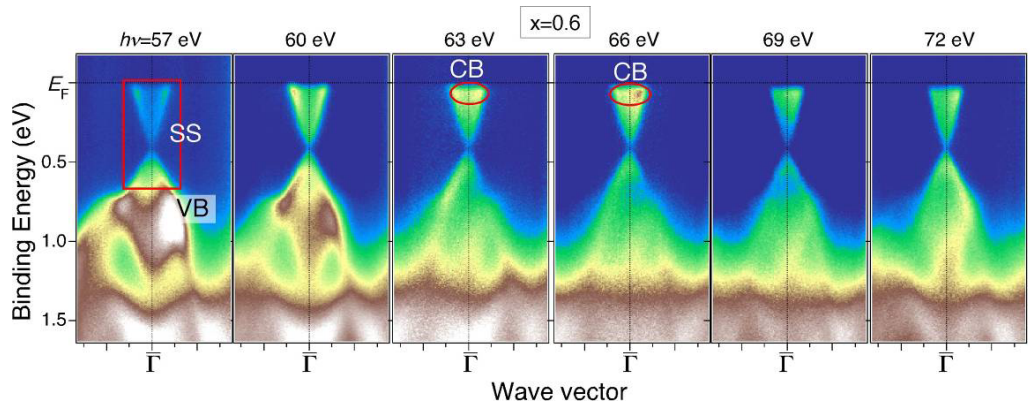

**Figure S1: Photon-energy dependence of the ARPES intensity for TlBi($S_{0.4}Se_{0.6}$)$_2$.** ARPES intensity plot near the Fermi level around the Brillouin-zone center as a function of wave vector and binding energy for the $x$=0.6 sample measured with various photon energies of synchrotron radiation.

We have confirmed that the observed band dispersion at $x$ = 0.6 showing a massive Dirac fermion behaviour indeed originates from the surface state by studying the photon-energy dependence. As shown in Fig. S1, the X-shaped dispersion (with the vanishing intensity at the crossing point) at the binding energy $E_B$ of 0 - 0.6 eV does not show any discernible change as a function of the photon energy $h\nu$. On the other hand, the bottom of the bulk conduction band (CB) located *within* the X-shaped band shows a significant change with the photon energy: it is invisible at $h\nu = 57$ eV, while it is well resolved at $h\nu = 63$ and 66 eV. Also, the bulk VB located at $E_B > 0.6$ eV shows a significant modulation with $h\nu$. This result is consistent with our identification of the origins of the bands. Most importantly, the Dirac gap in the surface band is robust against the photon energy variation, indicating its intrinsic character.

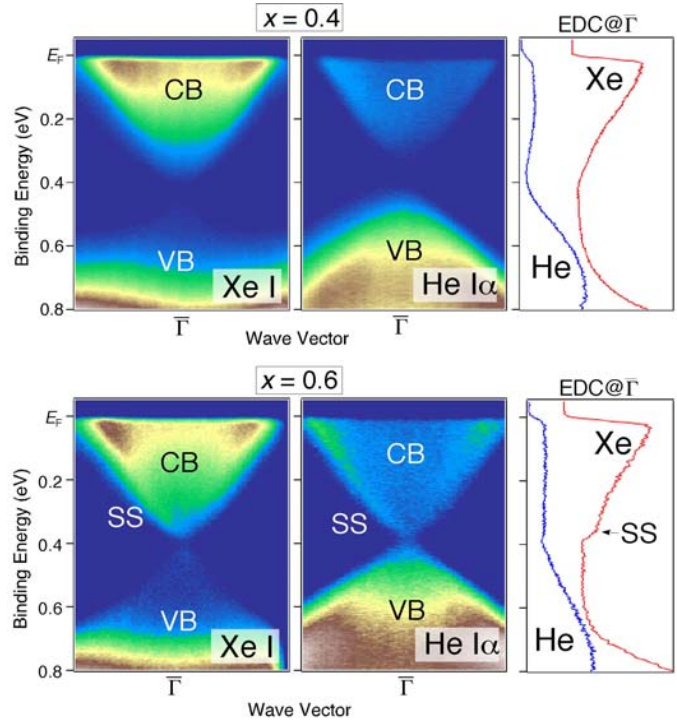

**Figure S2: Photon-energy dependence of the ARPES intensity for TlBi(S$_{1-x}$Se$_x$)$_2$ with $x$ = 0.4 and 0.6.** ARPES intensity plot near the Fermi level around the Brillouin-zone center as a function of wave vector and binding energy for the $x$=0.4 and 0.6 samples measured with the Xe I line ($h\nu$ = 8.437 eV) and the He I$\alpha$ line ($h\nu$ = 21.218 eV). Right panels display EDCs at the $\bar{\Gamma}$ point.

We have also confirmed the absence of the surface state at $x = 0.4$ and the appearance of it at $x = 0.6$ by changing the incident photon energy. Figure S2 shows a comparison of the ARPES intensity taken with $h\nu = 8.437$ and 21.218 eV for $x = 0.4$ (top) and $x = 0.6$ (bottom). The envelopes of the bands for $x = 0.4$ both above and below the gap clearly change with $h\nu$, indicating that those bands are of bulk origin and that there is no clear surface state. In contrast, at $x = 0.6$ we identify a surface band which can be clearly seen in the He I intensity. Indeed, energy distribution curves (EDCs) at the $\bar{\Gamma}$ point for $x = 0.6$ (right panel) show a sudden drop near the Dirac point (see arrow) corresponding to the bottom of the upper branch of the surface band, while such a discontinuous behavior is absent in $x = 0.4$. This result gives evidence that $x = 0.4$ belongs to the non-topological phase while $x = 0.6$ to the topological phase, supporting the occurrence of the quantum phase transition (QPT) at $x \sim 0.5$.

In addition, the data for $x = 0.4$ suggest that the bottom of the bulk CB is located only slightly above $E_B = 0.4$ eV, while the top of the bulk VB is only slightly below that, meaning that the bulk band gap is actually very small and that the estimate of the band gap based solely on the data with $h\nu = 8.437$ eV gives only the upper bound. Hence, we assigned due error bars to our estimate of the bulk band gap shown in Fig. 3**d** of the main text. As for the estimation of the bulk band gap for $x > 0.5$, we used synchrotron radiation sources to distinguish the surface and bulk bands by determining the three-dimensional band dispersion (see the caption of Figs. 3**c** and **d** in the text).

We note that the surface-band intensity was found to be enhanced by aging the sample surface in ultrahigh vacuum. During the ARPES experiment, we thus paid special attention to remove the aging effect, by measuring all the samples within 10 hours after cleaving during which we did not observe any detectable changes in the spectral feature.

## 3. Temperature and Momentum Dependence of the Surface State at $x = 0.6$.

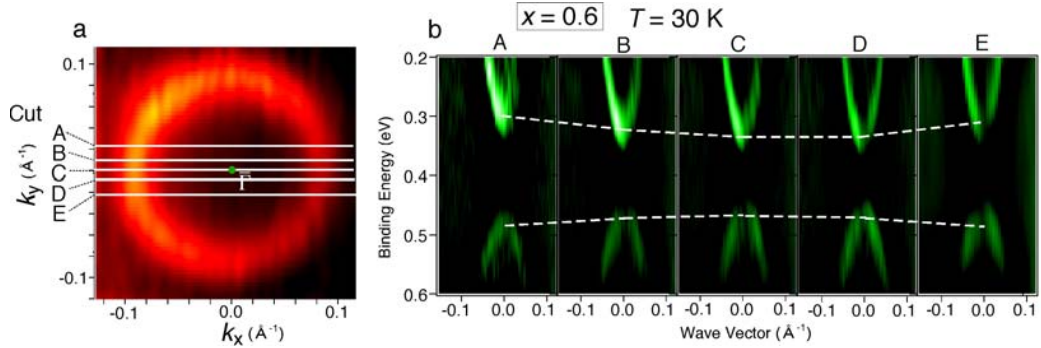

**Figure S3: Fermi surface and momentum dependence of the surface band dispersion for TlBi(S$_{1-x}$Se$_x$)$_2$ with $x = 0.6$. a**, ARPES intensity plot at the Fermi level as a function of two-dimensional wave vector. **b,** Second-derivative ARPES intensity for cuts A-E near the $\bar{\Gamma}$ point at $x = 0.6$. Both were measured with the Xe I line ($h\nu = 8.437$ eV). Dashed lines trace the top and bottom of surface bands.

In all the measured samples, we confirmed the location of the $\bar{\Gamma}$ point in the Brillouin zone, by mapping the Fermi surface in the two-dimensional momentum space and checking the surface band dispersion along several cuts near the $\bar{\Gamma}$ point. We display in Fig. S3 an example of such measurements for $x = 0.6$. Evidently, the lower and upper branches of the surface bands are well separated everywhere in the momentum space around the $\bar{\Gamma}$ point, and the momentum separation between the two bands becomes minimal exactly at the $\bar{\Gamma}$ point, demonstrating that the gap is not due to a misalignment of the measured sample. The estimated momentum accuracy for the determination of the $\bar{\Gamma}$ point is better than $0.005 \text{Å}^{-1}$.

To elucidate the character of the Dirac gap in more detail, we have performed a temperature-dependent ARPES measurement for $x = 0.6$ and show the result in Fig. S4. The Dirac gap, which is most clearly visible at 30 K, does not present strong temperature dependence, and it appears to persist even above 200 K. In fact, it might slightly increase upon heating the sample, though the intensity of the surface band significantly weakens at high temperature. These results strongly suggest that it is hard to explain the origin of the Dirac gap in terms of the formation of a magnetic order or magnetic transitions, which is consistent with our SQUID measurement (see Section 6 of this Supplementary Information). The larger apparent gap at higher temperature further implies that the Dirac gap is somehow related to fluctuations or randomness, since such effects at the surface might be thermally enhanced at higher temperatures.

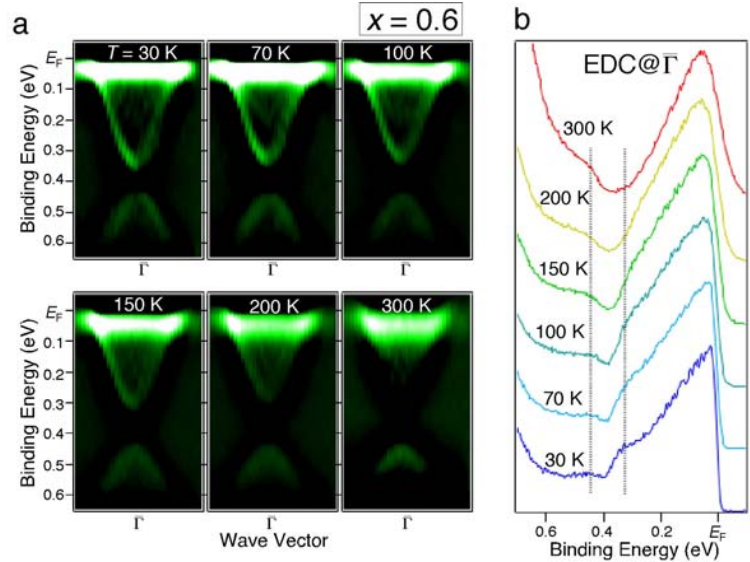

**Figure S4: Temperature dependence of the Dirac gap for TlBi(S$_{1-x}$Se$_x$)$_2$ with $x = 0.6$.** **a**, Temperature dependence of the second-derivative ARPES intensity through the $\bar{\Gamma}$ point for $x = 0.6$ as a function of wave vector and binding energy, measured with the Xe I line ($h\nu = 8.437$ eV). **b**, Temperature dependence of the EDC at the $\bar{\Gamma}$ point. Dashed lines are guides for the location of the lower and upper surface bands at 30 K.

## 4. Confirmation of the Homogeneity in the Crystals using EPMA

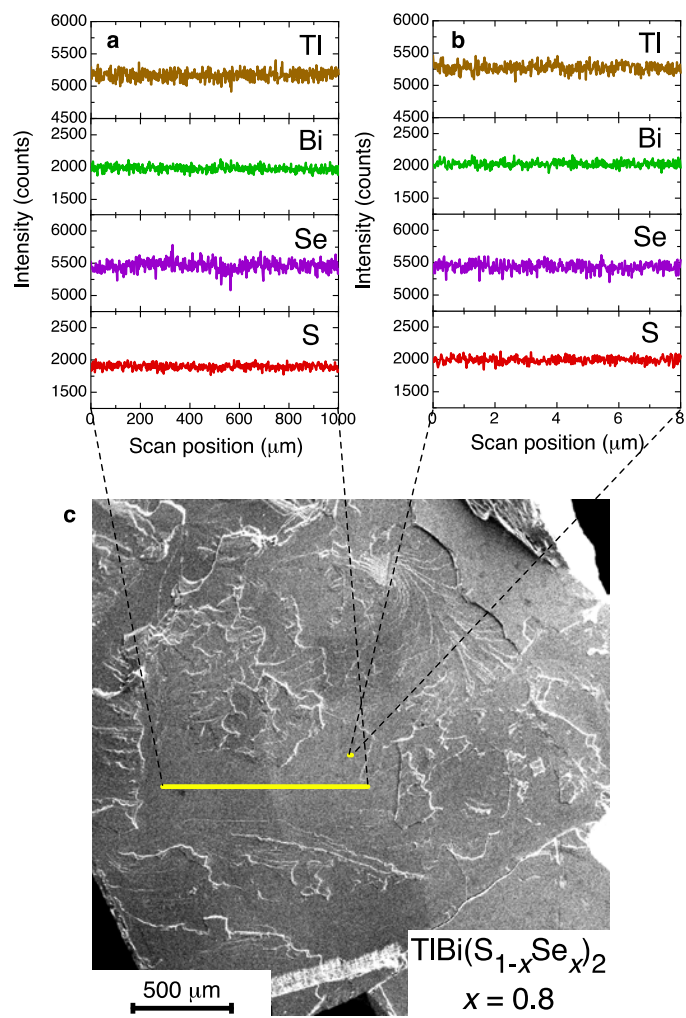

**Figure S5: Electron probe microanalysis of the homogeneity. a-b**, EPMA intensity profiles of line scans for the four elements. **c**, a SEM image of a single crystal of TlBi(S$_{1-x}$Se$_x$)$_2$ with $x = 0.8$. The scanned lines are shown in the image.

In order to confirm a homogeneous solution of selenium and sulfur in our single crystals, the electron probe microanalysis (EPMA) was performed. Figures S5**a** and S5**b** show intensity profiles of the line scans with different lengths for the four constituent elements in the TlBi(S$_{1-x}$Se$_x$)$_2$ crystal for $x = 0.8$. None of the scans show segregation of any of the elements. Figure S5**c** shows a scanning electron microscope (SEM) image in which the scanned lines are shown by solid lines. The length of the longer line corresponds to 1 mm and the shorter one is 8 µm.

## 5. X-Ray Diffraction Analysis

The lattice parameters in our TlBi($S_{1-x}Se_x$)$_2$ crystals were obtained by the x-ray diffraction (XRD) analysis for all the compositions studied using powered samples (Fig. S6**a**). Both of the lattice constants *a* and *c* systematically decrease with decreasing *x*. Figure S6**b** shows that the in-plane Bragg peaks used for determining the data in Fig. S6**a** show no noticeable broadening with *x*; also, Fig. S7 shows the (006) peak in the XRD data taken on single crystal samples, which present little broadening upon changing *x*. These data together indicate a high homogeneity of our crystals and good crystallinity in both the in-plane and the *c*-axis directions even in the solid-solution compositions. Note that all the samples used for the EPMA and XRD analyses were taken from the same batch as those used in the ARPES experiments.

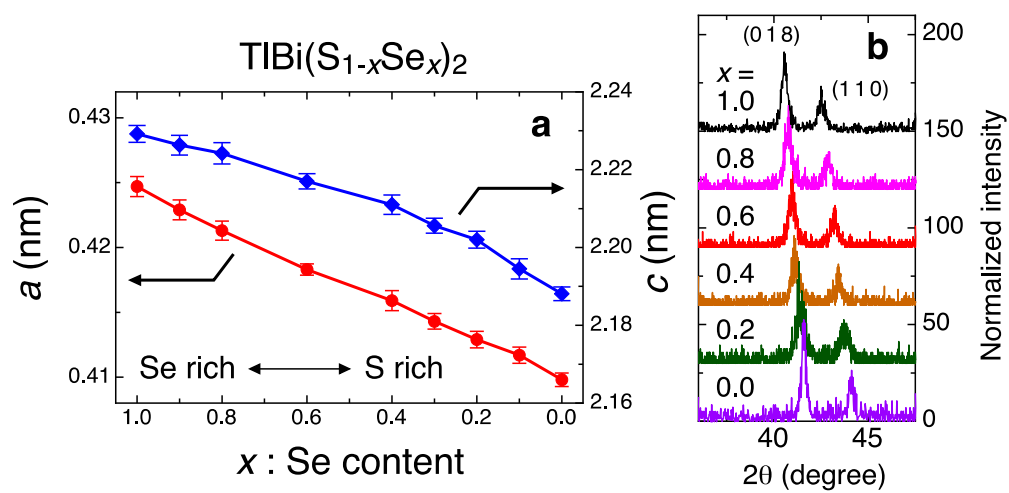

**Figure S6: a, Evolution of the lattice parameters.** Lattice constants *a* (filled red circles) and *c* (filled blue diamonds) determined from power XRD are plotted as functions of the Se content *x*. Error bars originate from the standard deviation of the XRD peak position in the fitting procedure. **b**, **Powder XRD peaks**. Normalized intensity profiles of the (018) and (110) peaks in the powdered TlBi($S_{1-x}Se_x$)$_2$ single crystals for a series of Se concentrations; the (018) peak reflects both the in-plane and the *c*-axis periodicities, whereas the (110) peak reflects solely the in-plane periodicities.

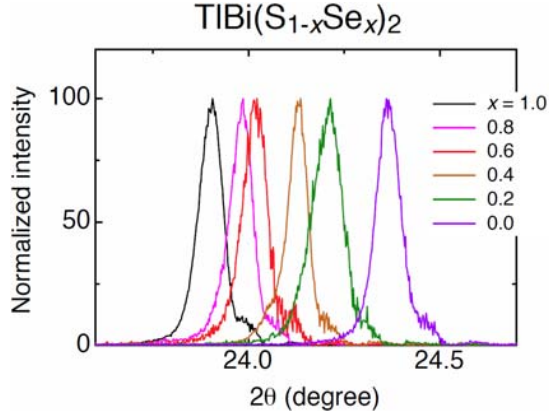

**Figure S7: Variation of the single crystal XRD peak.** Normalized intensity profiles of the (006) XRD peak in the TlBi($S_{1-x}Se_x$)$_2$ single crystals for a series of Se concentrations.

## 6. Magnetic Susceptibility

To check for the magnetic properties of TlBi($S_{1-x}Se_x$)$_2$, the temperature dependence of the magnetic susceptibility in a single crystal with $x = 0.8$ was measured with a commercial SQUID magnetometer (Quantum Design MPMS). The sample was found to be diamagnetic with the dc magnetic susceptibility of $-(3.9\pm0.2)\times10^{-7}$ emu/g, whose dependence on temperature was negligible below 300 K. This result indicates that there is no obvious magnetic transition taking place and that no detectable magnetic impurities, which might open the gap in the surface state, are present in this system.

The observed diamagnetism is characteristic of doped narrow-gap semiconductors. Indeed, the absolute value of the diamagnetic signal for $x = 0.8$ is comparable to that of $n$-type Bi$_2$Se$_3$ ($-3\times10^{-7}$ emu/g) (Ref. 1). Therefore, the bulk diamagnetism gives confidence in the narrow-gap band structure seen by ARPES on the surface.

## 7. Shubnikov-de Haas Oscillations

We were able to observe the Shubnikov-de Haas (SdH) oscillations for $x = 0.1$ and 0.4. Figure S8 shows the result of the SdH measurements in $x = 0.4$ at 1.5 K. The magnetic field was applied along the [111] direction (perpendicular to the cleavage plane). The oscillation frequency in this data is 157 T, which corresponds to the averaged Fermi wave vector $k_F = 0.0691$ Å$^{-1}$ for the extremal Fermi-surface cross section in the (111) plane. For $x = 0.1$, the frequency is 255 T corresponding $k_F = 0.0880$ Å$^{-1}$. Those bulk transport measurements of the Fermi surface are essentially consistent with the bulk band structure seen by ARPES, though the band-bending effect on the surface may have shifted the Fermi level in ARPES data to some extent.

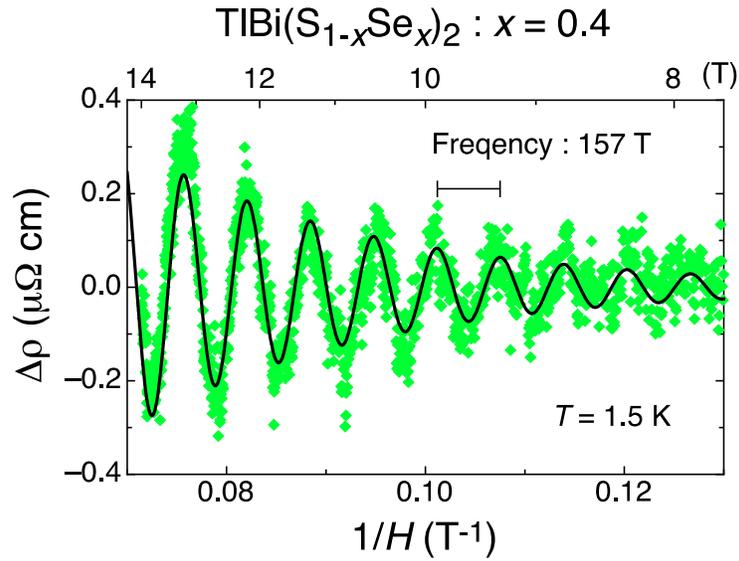

**Figure S8: Shubnikov-de Haas oscillations for *x* = 0.4.** The oscillatory component of the resistivity is plotted vs. 1/*H*. The background magnetoresistance was subtracted from the resistivity data. The solid line shows a fitting to the data, demonstrating that the oscillation is essentially composed of a single frequency of 157 T.

**Reference**

1. Kulbachinskii, V. A. *et al.* Low-temperature ferromagnetism in a new diluted magnetic semiconductor $Bi_{2-x}Fe_xTe_3$. *JETP Lett.* **73**, 352-356 (2001).